\documentclass[10pt]{article}
\usepackage{graphicx,epsf,bm}
\usepackage{amsmath,amsfonts,amssymb,amscd,bm}
\usepackage{flushend}
\usepackage[utf8]{inputenc}
\usepackage{tabularx}
\usepackage[usenames,dvipsnames]{color}
\usepackage{savesym}
\savesymbol{checkmark}
\usepackage{dingbat}
\usepackage{float}
\usepackage{bm}
\usepackage{placeins}
\usepackage{authblk}
\usepackage{rotating}
\usepackage{float}
\usepackage{tikz}
\usepackage{tikz-3dplot}
\usetikzlibrary{arrows}

\newcommand{\problemdomain}{\Omega}
\newcommand{\scatname}{\mathcal{S}}
\newcommand{\totname}{\mathcal{T}}
\newcommand{\scatdomain}{\problemdomain_{\scatname}}
\newcommand{\totaldomain}{\problemdomain_{\totname}}
\newcommand{\radcontrib}{r}
\newcommand{\scatcontrib}{s}
\newcommand{\primalgrid}{\mathcal{G}}
\newcommand{\dualgrid}{\mathcal{\tilde{G}}}

\newcommand{\normal}{\field{n}}
\newcommand{\field}[1]{\bm{#1}}

\newcommand{\tensor}[1]{\field{#1}}

\newcommand{\rot}{\nabla\times}

\newcommand{\primaledge}{e}
\newcommand{\primalface}{f}
\newcommand{\dualedge}{\tilde{e}}
\newcommand{\dualface}{\tilde{f}}

\newcommand{\primalcurl}{\mathbf{C}}
\newcommand{\dualcurl}{\mathbf{C}^T}
\newcommand{\numatrix}{\mathbf{M}_{\nu}}
\newcommand{\epsmatrix}{\mathbf{M}_{\epsilon}}
\newcommand{\ymatrix}{\mathbf{M}_{Y}}
\newcommand{\mmf}{\mathbf{F}}
\newcommand{\emf}{\mathbf{U}}
\newcommand{\magflux}{\mathbf{\Phi}}
\newcommand{\elecflux}{\mathbf{\Psi}}
\newcommand{\bordermmf}{{\mmf^b}}

\newcommand{\basepropagation}{(\dualcurl\numatrix\primalcurl - \omega^2\epsmatrix)\emf}
\newcommand{\admittancecontrib}{i\omega\ymatrix\emf}

\newcommand{\interface}{\Sigma}
\newcommand{\localeq}{v}

\usetikzlibrary{arrows,positioning, patterns}
\usetikzlibrary{decorations.markings}

\newcommand{\rotateRPY}[4][0/0/0]
{   \pgfmathsetmacro{\rollangle}{#2}
    \pgfmathsetmacro{\pitchangle}{#3}
    \pgfmathsetmacro{\yawangle}{#4}

    \pgfmathsetmacro{\newxx}{cos(\yawangle)*cos(\pitchangle)}
    \pgfmathsetmacro{\newxy}{sin(\yawangle)*cos(\pitchangle)}
    \pgfmathsetmacro{\newxz}{-sin(\pitchangle)}
    \path (\newxx,\newxy,\newxz);
    \pgfgetlastxy{\nxx}{\nxy};

    \pgfmathsetmacro{\newyx}{cos(\yawangle)*sin(\pitchangle)*sin(\rollangle)-sin(\yawangle)*cos(\rollangle)}
    \pgfmathsetmacro{\newyy}{sin(\yawangle)*sin(\pitchangle)*sin(\rollangle)+ cos(\yawangle)*cos(\rollangle)}
    \pgfmathsetmacro{\newyz}{cos(\pitchangle)*sin(\rollangle)}
    \path (\newyx,\newyy,\newyz);
    \pgfgetlastxy{\nyx}{\nyy};

    \pgfmathsetmacro{\newzx}{cos(\yawangle)*sin(\pitchangle)*cos(\rollangle)+ sin(\yawangle)*sin(\rollangle)}
    \pgfmathsetmacro{\newzy}{sin(\yawangle)*sin(\pitchangle)*cos(\rollangle)-cos(\yawangle)*sin(\rollangle)}
    \pgfmathsetmacro{\newzz}{cos(\pitchangle)*cos(\rollangle)}
    \path (\newzx,\newzy,\newzz);
    \pgfgetlastxy{\nzx}{\nzy};

    \foreach \x/\y/\z in {#1}
    {   \pgfmathsetmacro{\transformedx}{\x*\newxx+\y*\newyx+\z*\newzx}
        \pgfmathsetmacro{\transformedy}{\x*\newxy+\y*\newyy+\z*\newzy}
        \pgfmathsetmacro{\transformedz}{\x*\newxz+\y*\newyz+\z*\newzz}

    }
}

\newcommand{\block}[7]%
{
    \pgfmathsetmacro{\xa}{#1}
    \pgfmathsetmacro{\ya}{#2}
    \pgfmathsetmacro{\za}{#3}
    \pgfmathsetmacro{\xb}{#4}
    \pgfmathsetmacro{\yb}{#5}
    \pgfmathsetmacro{\zb}{#6}

    \coordinate (a) at (\xa, \ya, \za);
    \coordinate (b) at (\xb, \ya, \za);
    \coordinate (c) at (\xb, \yb, \za);
    \coordinate (d) at (\xa, \yb, \za);

    \coordinate (a1) at (\xa, \ya, \zb);
    \coordinate (b1) at (\xb, \ya, \zb);
    \coordinate (c1) at (\xb, \yb, \zb);
    \coordinate (d1) at (\xa, \yb, \zb);

    \filldraw [fill=lightgray, opacity=#7] (a)--(a1)--(b1)--(b)--(a);
    \filldraw [fill=lightgray, opacity=#7] (a)--(b)--(c)--(d)--(a);
    \filldraw [fill=lightgray, opacity=#7] (a1)--(b1)--(c1)--(d1)--(a1);
    \filldraw [fill=lightgray, opacity=#7] (c)--(c1)--(d1)--(d)--(c);
    \filldraw [fill=lightgray, opacity=#7] (a)--(a1)--(d1)--(d)--(a);
}
\newcommand{\cone}[3]%
{
    \pgfmathsetmacro{\xshift}{#1}
    \pgfmathsetmacro{\yshift}{#2}
    \pgfmathsetmacro{\zshift}{#3}

    \coordinate (a) at (3.95+\xshift,0.45+\yshift,0.45+\zshift);
    \coordinate (b) at (3.95+\xshift,0.45+\yshift,1.05+\zshift);
    \coordinate (c) at (3.95+\xshift,1.05+\yshift,1.05+\zshift);
    \coordinate (d) at (3.95+\xshift,1.05+\yshift,0.45+\zshift);

    \coordinate (a1) at (6.55+\xshift,0+\yshift,0+\zshift);
    \coordinate (b1) at (6.55+\xshift,0+\yshift,1.5+\zshift);
    \coordinate (c1) at (6.55+\xshift,1.5+\yshift,1.5+\zshift);
    \coordinate (d1) at (6.55+\xshift,1.5+\yshift,0+\zshift);

    \filldraw [fill=lightgray,opacity=0.8] (a1)--(b1)--(c1)--(d1)--(a1);
    \filldraw [fill=lightgray,opacity=0.8] (d)--(a)--(a1)--(d1)--(d);
    \filldraw [fill=lightgray,opacity=0.8] (a)--(b)--(b1)--(a1)--(a);
    \filldraw [fill=lightgray,opacity=0.8] (b)--(c)--(c1)--(b1)--(b);
    \filldraw [fill=lightgray,opacity=0.8] (c)--(d)--(d1)--(c1)--(c);
    \filldraw [fill=lightgray,opacity=0.8] (a)--(b)--(c)--(d)--(a);
}
\newcommand{\coneassembly}%
{
    \block{6.55}{0}{0}{7.25}{3}{3}{0.8}
    \cone{0}{0}{0}
    \cone{0}{0}{1.5}
    \cone{0}{1.5}{0}
    \cone{0}{1.5}{1.5}
}

\tikzset{RPY/.style={x={(\nxx,\nxy)},y={(\nyx,\nyy)},z={(\nzx,\nzy)}}}

\begin{document}
    \title{Modeling of anechoich chambers with equivalent materials and equivalent sources}%
    \author[1]{Silvano Chialina}
    \author[2]{Matteo Cicuttin}
    \author[3]{Lorenzo Codecasa}
    \author[1]{Giovanni Solari}
    \author[2]{Ruben Specogna}
    \author[2]{Francesco Trevisan}

    \affil[1]{\small Emilab s.r.l., Amaro (UD) I-33020, Italy}
    \affil[2]{\small Università degli Studi di Udine, Dipartimento di Ingegneria Elettrica, Gestionale e Meccanica, I-33100, Udine, Italy}
    \affil[3]{\small Politecnico di Milano, Dipartimento di Elettronica, Informazione e Bioingegneria, I-20133, Milano, Italy}

    \maketitle

\begin{abstract}
Numerical simulation of anechoic chambers is a hot topic since it can provide useful data about the performance of the EMC site. However, the mathematical nature of the problem, the physical dimensions of the simulated sites and the frequency ranges pose nontrivial challenges to the simulation. Computational requirements in particular will quickly become unmanageable if adequate techniques are not employed. In this work we describe a novel approach, based on equivalent elements, that enables the simulation of large chambers with modest computational resources. The method is then validated against real measurement results.   
\end{abstract}

\section{Introduction}
The performance of an EMC site is usually validated either with experimental approaches, like \cite{carobbi}, and numerical approaches, like \cite{sim-large-anec}. Both approaches are of great importance and subject of current research; moreover they are not exclusive but complementary, because each one can confirm the validity of the other. On the numerical side, simulation of anechoic chambers is a very attractive topic since it can provide useful data about the performance of an EMC site in short times and with reduced costs. However mathematical properties of the underlying problem, combined with the huge physical dimensions of the typical sites, make its numerical solution very challenging.

In this work, we describe a novel approach for the numerical simulation of entire anechoich chambers in the frequency domain. Instead of describing the geometry using a mesh with all the geometric details, we use equivalent models for the cone-ferrite assemblies and for radiating elements. This allows us to considerably reduce the number of mesh elements, and thus the computational resources required for the simulation, without degrading the accuracy of the results. In particular, our method allowed to simulate 558 different configurations of antenna positions and frequencies in a very short time on a common dual-Xeon workstation.

To validate the presented method, an experiment was devised: we set up a transmitter attached to a dipole inside a semi-anechoic chamber, with a measuring antenna three meters away. We then simulated the same setup with our methodology and compared the numerical results with the real field measurements.

In the following sections we first describe the experimental setup and then we set an \emph{uncertainty budget} on the measurements. Then the numerical part is presented: a brief introduction to the \emph{Discrete Geometric Approach (DGA)}, the numerical method employed to solve the frequency domain wave propagation problem, is given. The technique of substituting complex radiating elements with equivalent ones, which we will refer to as \emph{equivalent radiator method}, is then presented; theoretical aspects are discussed together with the applicability conditions of the technique. In a subsequent section, an equivalent dipole is validated by comparing the simulation with the field computed from the closed-form expression of a dipole. Finally, the modelling of the entire anechoic chamber is presented together with the comparison between the numerical results and the measurements.

\section{Experimental setup description}
\begin{figure}[ht]
    \centering
    \includegraphics[width=0.95\linewidth]{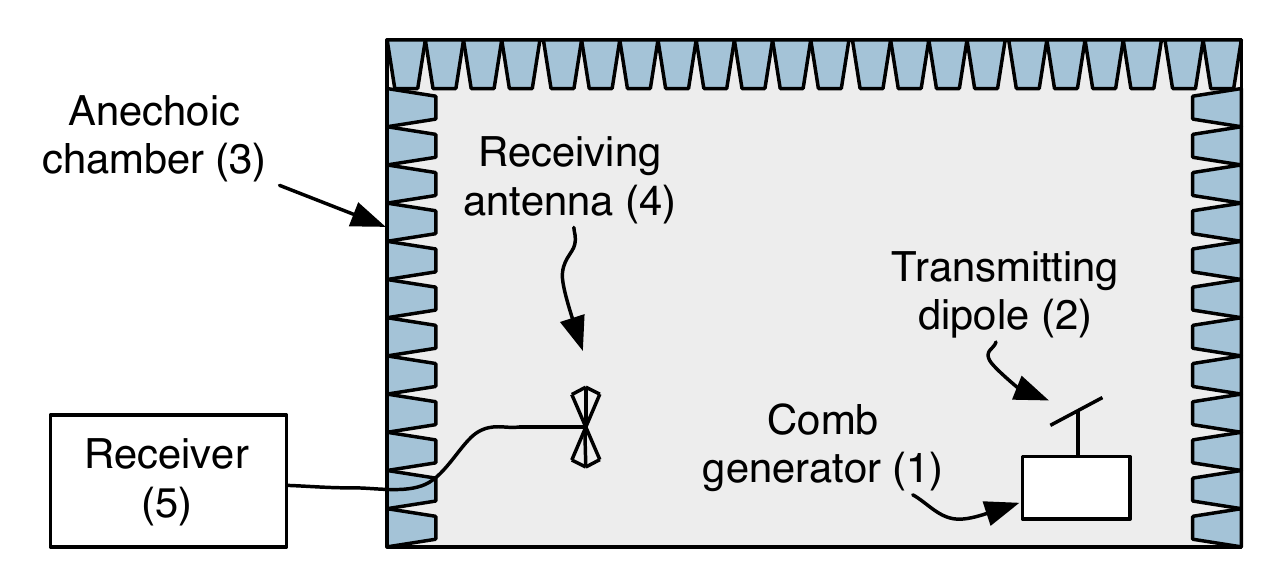}
    \caption{Setup used to measure the field radiated by the comb generator (1) attached to the test dipole (2) within the anechoic chamber (3) containing receiving antenna (4). Outside the chamber there is the EMI receiver (5).}
    \label{fig:setup1}
\end{figure}

Our experiment consisted in the measurement of the electric field produced by a RF radiator placed inside an anechoic chamber. The transmitter consisted in a comb generator connected to a specifically designed dipole. On the receiver side, a biconic antenna was attached to an EMI receiver (Fig. \ref{fig:setup1}). Transmitting and receiving antennae were placed at prescribed positions.

\subsection{Measurement setup}
Field measurements were made with transmitting and receiving antennas placed at heights of 1, 1.5 and 2 meters both in horizontal and vertical polarization, at a distance of three meters. This gave us a total of $3 \times 3 \times 2 = 18$ different measurement setups. Measurements consisted in sweeps from 90 to 390 MHz at steps of 10 MHz. The two frequency limits were chosen mostly for practical reasons. In particular, the antennae employed in the experiments cannot perform well outside that range. On the other hand, the chosen range allowed to use the same mesh for all the frequencies.

\subsection{Instrumentation}
The measurements involved a number of instruments, in particular:
\begin{itemize}
    \item Comb generator
    \item Hewlett-Packard 8591EM spectrum analyzer
    \item Agilent 9038A EMI receiver
    \item Ad-hoc transmitting dipole
    \item Schwarzbeck UBAA9114 biconical antenna
    \item Hewlett-Packard 8753E vector network analyzer (VNA)
\end{itemize}
The performance of all instruments was perfectly known, except for the comb generator and the trasmitting dipole, which required careful characterization (summarized in Fig.~\ref{fig:characterization}).
\subsubsection{The comb generator}
\begin{figure}[t]
    \centering
    \includegraphics[width=0.95\linewidth]{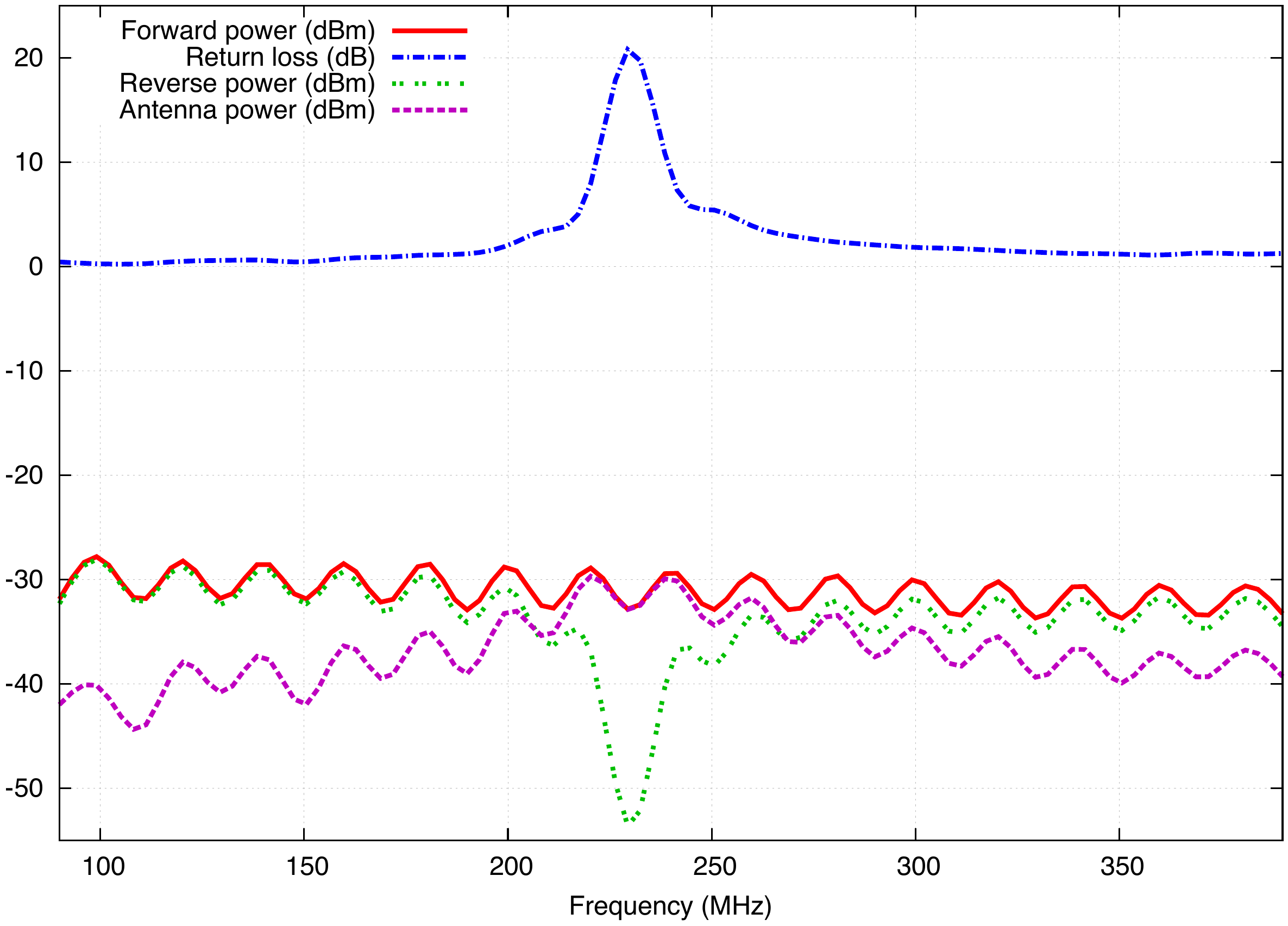}
    \caption{Summary of the data obtained from the characterization of the comb generator and the antenna: $P_{fwd}$, $P_{rev}$ and return loss and $P_{ant}$. The data is interpolated for clarity of visualization, but measurements were made at discrete steps of 10 MHz.}
    \label{fig:characterization}
\end{figure}

The comb generator is a specifically designed, battery-powered instrument capable to generate harmonics from 10 MHz to well above 1 GHz in steps of 10 MHz. Its power spectrum was measured to obtain, for each harmonic, the amount of \emph{forward power} $P_{fwd}$ supplied to the transmitting dipole. With the aid of a directional coupler, it was determined that the power spectrum of the comb generator changed by negligible amounts when changing its load. Moreover, as depicted in Fig.~\ref{fig:characterization}, the comb generator shows slightly different power levels for even and odd harmonics.
\subsubsection{Antenna characterization}
The characterization of the transmitting antenna was a central topic of the proposed scheme. The complex impedance $Z_{ant} = R_{ant} + iX_{ant}$ of the transmitting dipole was measured with the aid of the VNA. The impedance measurement was made for all the six antenna positions (three heights and two polarizations). From the impedance, return loss $RL$ was computed and used to calculate the \emph{reverse power} $P_{rev}$ flowing from the antenna back to the comb generator because of the mismatch. As a cross-check, the dipole was simulated with NEC2 code, which confirmed the measurements. The indirect measurement of $P_{rev}$ subsequently allowed to compute the \emph{net power} $P_{ant}$ with a good accuracy, despite of the heavy mismatch between the generator and the antenna. Finally, using $P_{ant}$ and the antenna impedance the supply current at the feedpoint of the dipole was computed as
\begin{equation}
    I_0 = \sqrt\frac{P_{ant}}{R_{ant}}. \label{eqn:dipole-current}
\end{equation}
The knowledge of $I_0$ will be needed in the numerical part of this work, as an input for the simulation. 

\subsection{Measurement chain uncertainty calculation}

\begin{sidewaystable}
    \centering
    \caption{Summary of the uncertainty contributes $x_i$ in the measurements.}
    \label{tbl:factors}
    \begin{tabular}{c|l|c|c|c|c|c|c}
        \parbox[t]{1.5cm}{\centering Symbol\\$(x_i)$\vspace{1mm}} & Meaning & Uncertainty & \parbox[t]{1.5cm}{\centering Probability\\distribution\vspace{1mm}} & \parbox[t]{1.5cm}{\centering Normal.\\factor\vspace{1mm}} & $u_i(x_i)$ & $c_i$ & $c_i u_i(x_i)$\\
        \hline
        \hline
        $R_i$ & Receiver reading & 0.1 & normal 1 & 1.00 & 0.10 & 1.00 & 0.10\\
        $L_{ar}$ & Receiver-antenna cable attenuation & 0.2 & normal 2 & 2.00 & 0.10 & 1.00 & 0.10\\
        $AF$ & UBAA9114 antenna factor & 0.5 & normal 2 & 2.00 & 0.25 & 1.00 & 0.25\\
        \hline
        \multicolumn{8}{c}{Receiver Corrections}\\
        \hline
        $V_{sw}$ & Sine wave voltage & 0.4 & normal 2 & 2.00 & 0.20 & 1.00 & 0.20\\
        $L_{mar}$ & Antenna-receiver mismatch & 0.1 & U-shape & 1.41 & 0.07 & 1.00 & 0.07\\

        \hline
        \multicolumn{8}{c}{Biconical antenna corrections}\\
        \hline

        $AF_i$ & AF frequency interpolation & 0.3 & rectangular & 1.73 &0.17 & 1.00 & 0.17\\
        $AF_h$ & AF height deviation & 1.5 & rectangular & 1.73 & 0.87 & 1.00 & 0.87\\
        $AF_{dir}$ & Directivity difference & 0.5 & rectangular & 1.73 & 0.29 & 1.00 & 0.29 \\

        \hline
        \multicolumn{8}{c}{Site corrections}\\
        \hline
        $dSA$ & Site imperfections (max) & 3.0 & triangular & 2.45 & 1.22 & 1.00 & 1.22 \\
        $dH$ & Table height & 0.1 & rectangular & 2.00 & 0.05 & 1.00 & 0.05 \\
        $Rr$ & Repeatability & 0.5 & normal 1 & 1.00 & 0.50 & 1.00 & 0.50 \\

        \hline
        \multicolumn{8}{c}{Transmitter corrections}\\
        \hline
        $TXi$ & Comb level & 0.5 & rectangular & 1.73 & 0.29 & 1.00 & 0.29 \\
        $dMta$ & Mismatch: antenna-comb generator & 0.24 & U-shape & 1.41 & 0.17 & 1.00 & 0.17 \\
        $Gtx$ & TX antenna gain & 2.0 & rectangular & 1.73 & 1.15 & 1.00 & 1.15  \\
        \hline
        \hline
        $u_t$ & \multicolumn{6}{l|}{Total uncertainty ($\sqrt{\sum_i (c_i u(x_i))^2}$)} & 2.03 \\
        $u_e$ & \multicolumn{6}{l|}{Expanded uncertainty (k=2) [dB]} & 4.05 \\

    \end{tabular}

\end{sidewaystable}

\begin{figure}[ht]
    \centering
    \includegraphics[width=0.75\linewidth]{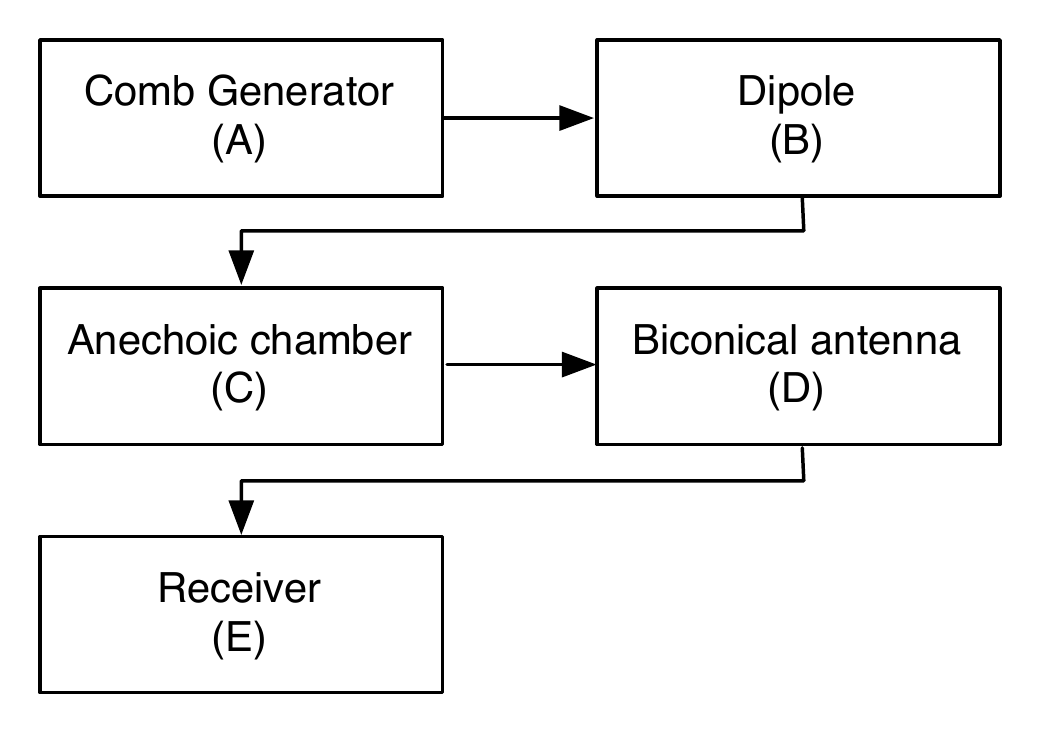}
    \caption{Schematic representation of the components involved in the measurement chain.}
    \label{fig:chain}
\end{figure}

To define a mathematical model of the measurement process, the entire measurement chain can be schematically illustrated as in Fig.~\ref{fig:chain}.
From the already mentioned comb generator, marked as (A), the signal goes to the dipole (B), which radiates the electromagnetic field in the semi-anechoic chamber (C). The semi-anechoic chamber is an EMC site that mimics the \emph{Open Area Test Site} (OATS in the following), whose primary characteristic is a metallic ground plane which behaves as a reflective surface. The mirror effect implies that on the receiving antenna the signal is the sum of a direct wave and a reflected wave which, in turn, produces different signal levels at different receiving antenna positions. Moreover, in a real anechoic chamber, there are multiple secondary signal paths due to the reflections on the walls and on the ceiling. These effects are considered as site imperfections and are taken into account in the \emph{Normalized Site Attenuation}, a parameter measured using standard procedures \cite{cispr-16-4-2}. Because of these effects, the field measurement at fixed positions, as we made in our experiments, is particularly challenging. At the chosen measurement points, the signal is received with a biconical antenna (D) and transferred to the receiver (E), which gives the experimental reading.
Each block depicted in Fig.~\ref{fig:chain} contributes to the measurement uncertainty, in particular the measured amplitude of the electric field $|\field{e}_m|$
\begin{equation}
    |\field{e}_m| = \sum_i x_i
\end{equation}
is given by the sum of the factors $x_i$ expressed in logarithmic scale, which are reported in Table~\ref{tbl:factors}. While some of them have a quite intuitive meaning (reported in Table~\ref{tbl:factors}) others deserve special attention, in particular:
\begin{itemize}
    \item Uncertainty on antenna factor $AF$: since the $AF$ is measured on a discrete set of frequency, there is an error due to interpolation, denoted as $AF_i$. Moreover, the antenna factor changes with antenna height and antenna axis direction, these effects are denoted by $AF_{h}$ and $AF_{dir}$.
    \item Site imperfections $dSA$: the site imperfections are accounted for by the \emph{Normalized Site Attenuation} \cite{cispr-16-4-2}. 
\end{itemize}
Moreover, some other important parameters were set to zero and thus omitted from Table~\ref{tbl:factors}, in particular:
\begin{itemize}
    \item Noise floor $V_{nf}$: The uncertainty due to the noise floor was considered negligible since all the measurements had an adequate signal-to-noise ratio ($> 10dB$).
    \item Cross-polarization $A_{cp}$ and unbalance $A_{bal}$: these two contributes are set to zero because of antenna specifications.
\end{itemize}

Once all the factors and their uncertainties are known, the total uncertainty $u_t$ and the extended uncertainty $u_e = 2u_t$ are calculated as \cite{gum1}
\begin{equation}
    u_t = \sqrt{\sum_i \left[c_i u(x_i)\right]^2},
\end{equation}
where $c_i$ is the sensitivity coefficient of the $i$-th contribute $x_i$, while $u_i(x_i)$ is its uncertainty value. The sensitivity coefficient is calculated as \cite{gum1}
\begin{equation}
    c_i = \frac{\partial |\field{e}_m|}{\partial x_i}.
\end{equation}

\section{Overview}
\subsection{The electromagnetic problem}
The electromagnetic wave propagation problem in the frequency domain at angular frequency $\omega$ is written as
\begin{align}
    \rot (\tensor{\nu} \rot \field{e}) - \omega^2\tensor{\epsilon}\field{e} = \mathbf{0}, \label{eqn:waveprop}
\end{align}
where $\field{e}$ is a complex-valued vector function of the position, representing the electric field while $\tensor{\epsilon}$ and $\tensor{\nu}$ are respectively the electric and the magnetic material positive definite tensors.
Problem (\ref{eqn:waveprop}) can be solved subject to specific boundary conditions, for example the admittance boundary condition
\begin{equation}
    \field{h} \times \normal = Y((\normal \times \field{e}) \times \normal), \label{eqn:cont-bnd-admit}
\end{equation}
where $Y = 1/Z$. Admittance is used since it arises naturally from the discrete formulation \cite{dga-admittance}.

\subsection{DGA formulation of the electromagnetic problem}
The DGA requires a discretization of the region $\problemdomain$ in which the problem is defined, consisting in a pair of interlocked grids $\primalgrid$ and $\dualgrid$. $\primalgrid$ is a tetrahedral grid while $\dualgrid$ is obtained from $\primalgrid$ by barycentric subdivision \cite{dga-admittance}. The electromagnetic quantities are defined on the geometric elements composing these grids, in particular
\begin{itemize}
    \item the electromotive force $U_i = \int_{\primaledge_i}\field{e} \cdot d\bm{l}$ on the primal edges $\primaledge_i \in \primalgrid$,
    \item the magnetic flux $\Phi_{i} = \int_{\primalface_i}\field{b} \cdot d\bm{s}$ on the primal faces $\primalface_i \in \primalgrid$,
    \item the magnetomotive force $F_i = \int_{\dualedge_i}\field{h} \cdot d\bm{l}$, on the dual edges $\dualedge_i \in \dualgrid$,
    \item the electric flux $\Psi_{i} = \int_{\dualface_i}\field{d} \cdot d\bm{s}$, on the dual faces $\dualface_i \in \dualgrid$.
    \end{itemize}
    Moreover, the standard face-edge incidence matrices $\primalcurl$ on $\primalgrid$ and $\dualcurl$ on $\dualgrid$ are introduced \cite{dga-admittance}. According to the DGA, the electromagnetic propagation problem (\ref{eqn:waveprop}) is written as
%
\begin{align}
    \basepropagation = \mathbf{0} \label{eqn:propagation-DGA},
\end{align}
where $\emf$ is an array collecting circulations of the electric field on primal edges, while symmetric positive definite material matrices $\epsmatrix$ and $\numatrix$ are calculated according to \cite{dga-matrices}.

\subsection{Impedance boundary condition}
The discrete counterpart of the admittance boundary condition (\ref{eqn:cont-bnd-admit}) is encoded as
\begin{equation}
    \bordermmf = \ymatrix\emf, \label{eqn:boundary-DGA}
\end{equation}
where the array $\bordermmf$ has the same number of elements of $\emf$, but its nonzero entries are only the ones associated to dual boundary edges $\dualedge_b$ of $\partial\problemdomain$ \cite{dga-admittance}, which are in one to one correspondence to primal boundary edges $\primaledge_b$.

The matrix $\ymatrix$ has also nonzero entries only in correspondence of the boundary edges. Its entries are calculated from the admittance parameter $Y$.
According to \cite{dga-admittance}, condition (\ref{eqn:boundary-DGA}) is integrated into (\ref{eqn:propagation-DGA}) as 
\begin{equation}
    \basepropagation + \admittancecontrib = \mathbf{0} \label{eqn:propagation-DGA-fb}. 
\end{equation}
%
%
%
\subsection{Plane wave excitation}
The impedance boundary condition can be extended to represent a plane wave excitation \cite{port-boundary-condition}, which is encoded in the right side of (\ref{eqn:propagation-DGA-fb}) as
\begin{align}
    \basepropagation &+ \admittancecontrib = -2i\omega\bordermmf^-, \label{eqn:complete-probl}
\end{align}
where $\bordermmf^-$ is nonzero only in correspondence of the edges of the boundary where the plane wave is forced and collects the magnetomotive forces due to the excitation.
\subsection{Unit cell model}
\begin{figure}[ht]
    \centering
    \includegraphics[width=0.65\linewidth]{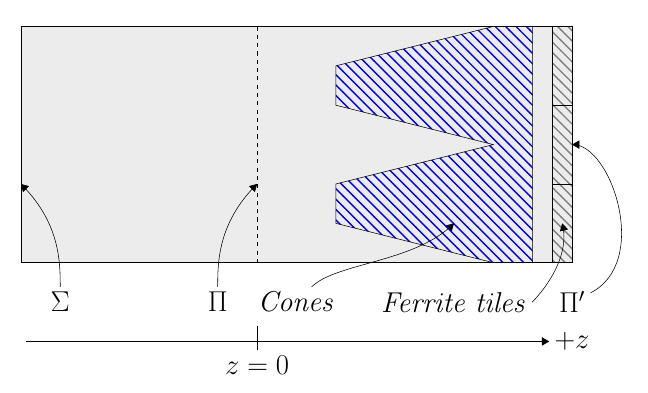}
    \caption{Cross section of the unit cell, showing pyramidal absorbers (cones) and ferrite tiles. Solid gray regions represent air.}
    \label{fig:uc-section}
\end{figure}

All the ideas exposed so far can be used to study the basic element of an anechoic wall, which we call \emph{unit cell} \cite{port-boundary-condition}. An unit cell is a section of an anechoic wall composed by $2\times 2$ absorbing cones and $3 \times 3$ ferrite tiles. In front of the cones and between the cones and the ferrite tiles there is air, as detailed in Fig.~\ref{fig:uc-section}.

The goal of studying the unit cell is to obtain an equivalent representation \cite{port-boundary-condition} in terms of an admittance boundary condition. This is done by the following steps:
\begin{itemize}
    \item Apply a plane wave excitation on the plane $\Sigma$
    \item Calculate wave impedance on the plane $\Pi$
    \item Remove all the materials inside the unit cell and translate the impedance calculated on $\Pi$ to $\Pi'$.
    \item Use the translated impedance as impedance boundary condition for entire walls in the simulation of the entire anechoic chamber.
\end{itemize}

\begin{figure}[ht]
    \centering
    \includegraphics[width=0.65\linewidth]{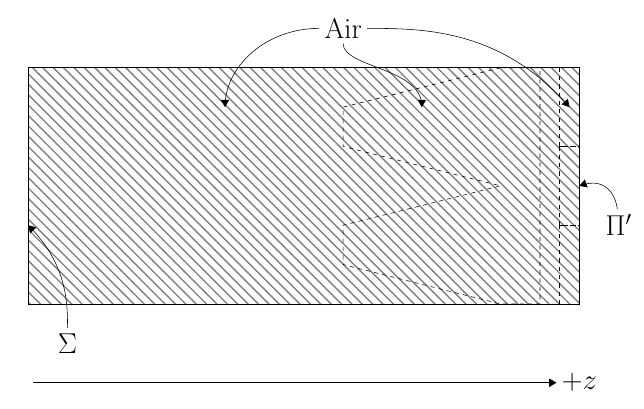}
    \caption{Cross section of the unit cell, with pyramidal absorbers and ferrite tiles removed (dashed lines are for reference). All the regions previously filled with absorbing material are now empty space filled with air.}
    \label{fig:uc-equiv}
\end{figure}

\section{Equivalent radiating elements}
%
%
In this section it will be shown how a generic radiating element, for example an antenna, can be substituted by an equivalent model. The idea is to substitute an arbitrarily complex object with a sphere radiating a field with the same characteristics of the original one. To achieve this goal, the domain $\problemdomain$ is partitioned in two regions $\scatdomain$ and $\totaldomain$.

\begin{figure}[ht]
    \centering
    \includegraphics[width=0.8\linewidth]{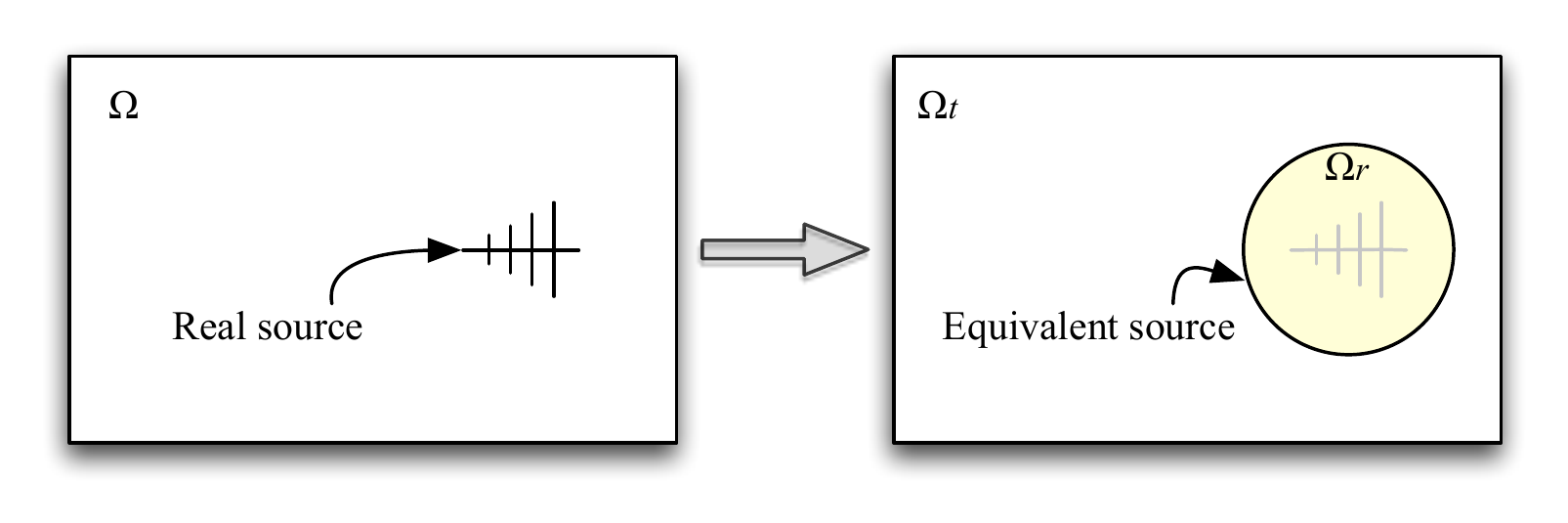}
    \caption{The real source (left) is transformed in an empty sphere (right). Because the field of the original source is projected onto the surface of the sphere, it radiates a field equivalent to the original one. }
    \label{fig:realToEquiv}
\end{figure}

The region $\scatdomain$ is the one that contains the radiator, while the region $\totaldomain$ is the remaining part. Moreover, in the region $\totaldomain$ the \emph{total field} is calculated, while in the region $\scatdomain$ only the \emph{scattering field} is computed (Fig. \ref{fig:realToEquiv}). In this way it is possible to evaluate the reaction of the environment to the field radiated by the radiator itself. The separation of $\totaldomain$ and $\scatdomain$ is obtained by introducing a \emph{boundary dual grid}, whose role is adjusting the discrete Maxwell's equations on the interface $\interface$ between the two regions $\totaldomain$ and $\scatdomain$. The adjustment consists in projecting on $\interface$ the fields due to the radiation of the original source: in the following sections it will be shown how this can be done \emph{locally} (i.e. element-wise). Global equations are then obtained by \emph{assembling} the local contributes in the usual way. In the following discussion local quantities will be denoted with the superscript $v$, except for the matrix $\primalcurl$, which we assume the local one unless otherwise noted.
Moreover, we choose that the edges of $\interface$ belong to $\scatdomain$. This means that tetrahedra in $\scatdomain$ and touching $\interface$ are the only affected by the modified Maxwell equations shown below. Every other tetrahedra in $\problemdomain$ is treated as usual.


\subsection{Ampère--Maxwell law}
\begin{figure}[h]
    \centering
    \includegraphics[width=0.9\linewidth]{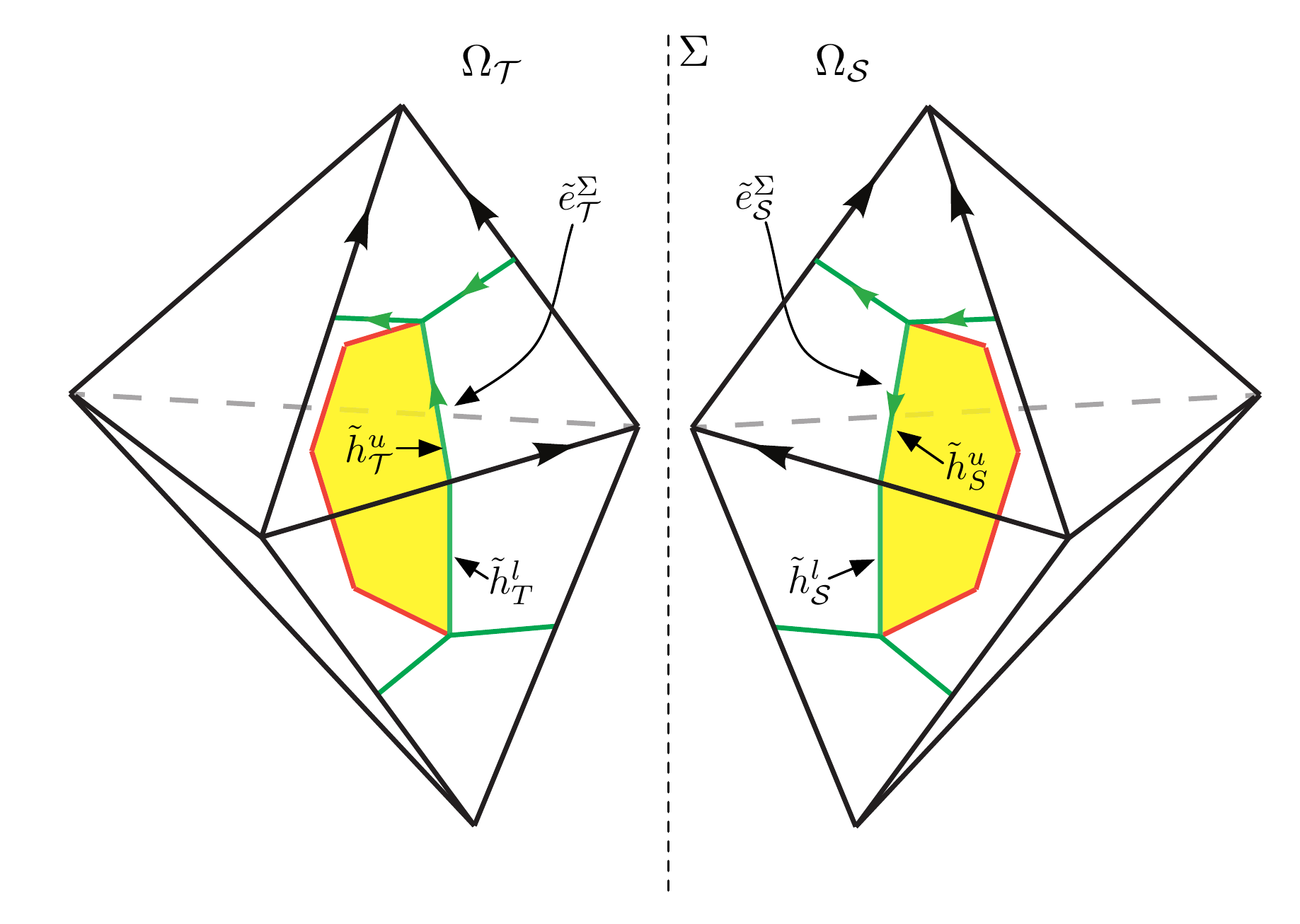}
    \caption{A cluster of tetrahedra having $\interface$ in common where \emph{boundary dual edges} are shown. Since half dual edges $\tilde{h}_{\totname}^{u}$ and $\tilde{h}_{\scatname}^{u}$ are the same edge but with opposite orientation, the associated magnetomotive forces $F_{\totname}$ and $F_{\scatname}$ sum to zero. But since $F_{\scatname}$ = $F_{\scatname\scatcontrib}$ + $F_{\scatname\radcontrib}$, also $F_{\totname}$~+~$F_{\scatname\scatcontrib}$~=~-$F_{\scatname\radcontrib}$ holds.}
    \label{fig:circulation}
\end{figure}

Consider, in relation to Fig. \ref{fig:circulation}, the boundary dual edges $\tilde{e}_{\totname}^{\interface}$ and $\tilde{e}_{\scatname}^{\interface}$, each spanning two tetrahedra. Both $\tilde{e}_{\totname}^{\interface}$ and $\tilde{e}_{\scatname}^{\interface}$ can be split in two parts, which we call \emph{half edges}, such that $\tilde{e}_{\totname}^{\interface} = \tilde{h}_{\totname}^{u} \cup \tilde{h}_{\totname}^{l}$ and $\tilde{e}_{\scatname}^{\interface} = \tilde{h}_{\scatname}^{u} \cup \tilde{h}_{\scatname}^{l}$. Each half edge belongs to a single tetrahedron, for example $\tilde{h}_{\totname}^{u}$ belongs to the upper tetrahedron in $\problemdomain_{\totname}$ while $\tilde{h}_{\scatname}^{l}$ belongs to the lower tetrahedron in $\problemdomain_{\scatname}$ (Fig. \ref{fig:circulation}). To write the contribute to the Ampère--Maxwell law for the single tetrahedron $v$, only the half edges belonging to $v$ must be considered. Reasoning on the two upper tetrahedra in Fig. \ref{fig:circulation}, Ampère--Maxwell law is obtained by observing that the magnetomotive force $F_{\totname}$ on $\tilde{h}_{\totname}^{u}$ and $F_{\scatname}$ on $\tilde{h}_{\scatname}^{u}$ must satisfy the relation $F_{\totname}^{\interface} + F_{\scatname}^{\interface} = 0$ because $\tilde{h}_{\totname}^{u}$ and $\tilde{h}_{\scatname}^{u}$ are in fact the same edge but with opposite orientation. 
However, since we impose the excitation on the scattering subdomain, the magnetomotive force $F_{\scatname}$ can be further decomposed in the unknown scattered contribute $F_{\scatname\scatcontrib}$ and in the known radiated contribute $F_{\scatname\radcontrib}$. This implies that the balance of the magnetomotive forces must satisfy the condition $F_{\totname}$~+~$F_{\scatname\scatcontrib}$~=~-$F_{\scatname\radcontrib}$. Thus, the local Ampère--Maxwell law for a tetrahedron $v$ in $\problemdomain_{\scatname}$ touching $\interface$ is written as
\begin{equation}
    \dualcurl\mmf^{\localeq} - \mmf_\radcontrib^{\localeq} = i\omega\mathbf\elecflux^{\localeq}, \label{eqn:locam}
\end{equation}
where the term $\mmf_\radcontrib^{\localeq}$ collects the magnetomotive forces on the half edges of $v$ due to the excitation.

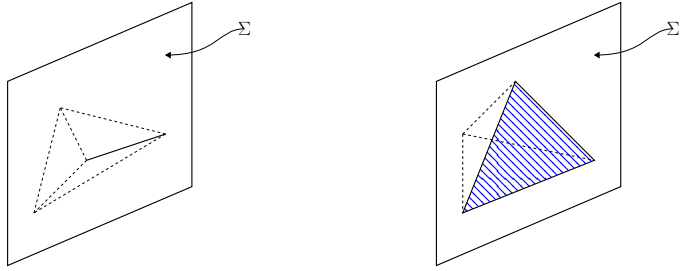
\begin{figure}[ht]
    \centering
    \hfill%
    \begin{minipage}{0.40\linewidth}
        \scalebox{0.35}{
\begin{tikzpicture}

\coordinate (v1) at (-10,4);
\coordinate (v2) at (-3,7);
\coordinate (v3) at (-10,-3);
\coordinate (v4) at (-3,0);

\draw  [very thick ](v1) -- (v3) -- (v4) -- (v2) -- cycle ;

\coordinate (te1) at (-7,1);
\coordinate (te2) at (-4,2);
\coordinate (te3) at (-9,-1) {};
\coordinate (te4) at (-8,3) {};
\draw [very thick] (te1) edge (te2);
\draw [dashed] (te1) edge (te3);
\draw [dashed] (te1) edge (te4);
\draw [dashed] (te2) edge (te3);
\draw [dashed] (te2) edge (te4);
\draw [dashed] (te3) edge (te4);

\node at (-4,5) {};
\node at (-1,6) {\huge $\Sigma$};

\draw [-triangle 60](-1,6) .. controls (-2,6) and (-2,5) .. (-4,5);
\end{tikzpicture}
                    }
    \end{minipage}%
    \hfill%
    \begin{minipage}{0.45\linewidth}
        \scalebox{0.35}{

\begin{tikzpicture}

\coordinate (v1) at (-10,4);
\coordinate (v2) at (-3,7);
\coordinate (v3) at (-10,-3);
\coordinate (v4) at (-3,0);

\draw  [very thick ](v1) -- (v3) -- (v4) -- (v2) -- cycle ;

\coordinate (te1) at (-9,-1) {};
\coordinate (te2) at (-4,1);
\coordinate (te3) at (-9,2) {} {};
\coordinate (te4) at (-7,4) {} {};
\draw [pattern=north west lines, pattern color=blue, very thick] (te1) -- (te2) -- (te4) -- cycle;

\node at (-4,5) {};
\node at (-1,6) {\huge $\Sigma$};

\draw [-triangle 60](-1,6) .. controls (-2,6) and (-2,5) .. (-4,5);

\draw [dashed] (te3) edge (te1);
\draw [dashed] (te3) edge (te4);
\draw [dashed] (te3) edge (te2);
\end{tikzpicture}

        }
    \end{minipage}
    \caption{The two cases of interest of elements touching $\interface$: the \emph{edge} case on the left side and the \emph{face} case on the right side.}
    \label{fig:cases}
\end{figure}

\subsection{Faraday--Neumann law}
To derive the expression for the Faraday--Neumann law, tetrahedra on the interface $\interface$ must be studied. In particular, two cases are identified:
\begin{itemize}
    \item Volume element with an edge lying on $\interface$
    \item Volume element with a face lying on $\interface$
\end{itemize}

%
\subsubsection{The case of a face on $\interface$}
Consider a volume element $v$ lying on $\interface$, the faces of which are $f_1, \ldots, f_4$ and their magnetic fluxes are collected in the array $\magflux^{\localeq}~=~(\phi_1^v,\ldots,\phi_4^v)$. Assume, without loss of generality, that the face on $\interface$ is $f_1$, so $v \cap \interface = f_1$ (Fig. \ref{fig:cases}). Assume also that the edges surrounding $f_1$ are $e_1, e_2, e_3$. Both electromotive forces $U_1^v, U_2^v, U_3^v$ and flux $\phi_1^v$ on $f_1$ can be decomposed in an unknown scattering component and in a known radiated component, obtaining $U_k^v~=~U_{k,\scatcontrib}^v~+~U_{k,\radcontrib}^v$ with $k \in \{1,2,3\}$ and $\phi_1^v~=~\phi_{1\scatcontrib}^v~+~\phi_{1\radcontrib}^v$. In this case both sides of the Faraday--Neumann law are modified, as follows:
\begin{equation}
    \primalcurl
    \begin{pmatrix}
        U_{1\scatcontrib}^v + U_{1\radcontrib}^v\\
        U_{2\scatcontrib}^v + U_{2\radcontrib}^v\\
        U_{3\scatcontrib}^v + U_{3\radcontrib}^v\\
        U_4^v\\
        U_5^v\\
        U_6^v
    \end{pmatrix} = -i\omega
    \begin{pmatrix}
        \phi_{1\scatcontrib}^v + \phi_{1\radcontrib}^v\\
        \phi_2^v\\
        \phi_3^v\\
        \phi_4^v
\end{pmatrix}.
\end{equation}

Separating known and unknown quantities and writing the equation in compact form we obtain
\begin{equation}
    \primalcurl \left( \emf^{\localeq}_{\radcontrib} + \emf^{\localeq}_{\scatcontrib} \right) = -i\omega\left( \magflux^{\localeq}_{\radcontrib} + \magflux^{\localeq}_{\scatcontrib} \right). \label{eqn:locfn}
\end{equation}

\subsubsection{The case of an edge on $\interface$}
In this case (\ref{eqn:locfn}) remains valid by setting to zero the radiated contributes of the geometric elements not touching $\interface$ (Fig.\ref{fig:cases}).
\subsection{Constitutive relations}
Almost the same reasoning carried out for the Faraday--Neumann law must be made for the constitutive relations when dealing with volume elements lying on $\interface$. There are again two cases, the one of the edge on $\interface$ and the one of the face on $\interface$.
%
%

\subsubsection{The case of a face on $\interface$}
In this case both electromotive forces and magnetic fluxes are split in radiated and scattered contributes, obtaining
\begin{equation}
\mathbf{\elecflux}^{\localeq}=\epsmatrix^{\localeq}
\begin{pmatrix}
    U_{1\scatcontrib}^v + U_{1\radcontrib}^v\\
    U_{2\scatcontrib}^v + U_{2\radcontrib}^v\\
    U_{3\scatcontrib}^v + U_{3\radcontrib}^v\\
    U_4^v\\
    U_5^v\\
    U_6^v
\end{pmatrix} = 
\epsmatrix^{\localeq} \begin{pmatrix}U_{1\scatcontrib}^v\\U_{2\scatcontrib}^v\\U_{3\scatcontrib}^v\\U_4^v\\U_5^v\\U_6^v\end{pmatrix} + 
\epsmatrix^{\localeq} \begin{pmatrix}U_{1\radcontrib}^v\\U_{2\radcontrib}^v\\U_{3\radcontrib}^v\\0\\0\\0\end{pmatrix},
\end{equation}

\begin{equation}
\mmf^{\localeq}=\numatrix^{\localeq}
\begin{pmatrix}
    \phi_{1\scatcontrib}^v + \phi_{1\radcontrib}^v\\
    \phi_2^v\\
    \phi_3^v\\
    \phi_4^v
\end{pmatrix} = 
\numatrix^{\localeq} \begin{pmatrix}\phi_{1\scatcontrib}^v\\\phi_2^v\\\phi_3^v\\\phi_4^v\end{pmatrix} + 
\numatrix^{\localeq} \begin{pmatrix}\phi_{1\radcontrib}^v\\0\\0\\0\end{pmatrix},
\end{equation}

or, written in compact form,

\begin{align}
    \elecflux^{\localeq} &= \epsmatrix^{\localeq}(\emf_{\scatcontrib}^{\localeq} + \emf_{\radcontrib}^{\localeq}), \label{eqn:locce}\\
    \mmf^{\localeq} &= \numatrix^{\localeq} (\magflux_{\scatcontrib}^{\localeq} + \magflux_{\radcontrib}^{\localeq})\label{eqn:loccm}.
\end{align}

\subsubsection{The case of an edge on $\interface$}
As in the case of the Faraday--Neumann law, in this case there are no magnetic fluxes across $\interface$ and only one voltage along an edge on $\interface$. Constitutive relations (\ref{eqn:locce}) and (\ref{eqn:loccm}) continue to remain valid by setting to zero the radiated contribute of the entries corresponding to geometric entities not on $\interface$ (Fig. \ref{fig:cases}).

\subsection{From local equations to the global equation}
Until now we reasoned in terms of single mesh volumes, so an assembly phase must be carried out in order to obtain the global equation. Consider (\ref{eqn:locam}), (\ref{eqn:locfn}), (\ref{eqn:locce}) and (\ref{eqn:loccm}): by solving (\ref{eqn:locam}) for $(\magflux^{\localeq}_{\radcontrib} + \magflux^{\localeq}_{\scatcontrib})$ then substituting (\ref{eqn:locce}), (\ref{eqn:locam}), (\ref{eqn:loccm}) and rearranging, the expression
\begin{equation}
    \mathbf{K}^{\localeq}\emf^{\localeq}_{\scatcontrib} = -\mathbf{K}^{\localeq}\emf^{\localeq}_{\radcontrib} - i\omega\mmf^{\localeq}_{\radcontrib} \label{eqn:localprop}
\end{equation}
is obtained, where $\mathbf{K}^{\localeq} = \dualcurl\numatrix^{\localeq}\primalcurl - \omega^2\epsmatrix^{\localeq}$. Assembling element by element in the usual way, the equation
\begin{equation}
    \mathbf{K}\emf = -\mathbf{K}\emf_{\radcontrib} - i\omega\mmf_{\radcontrib} \label{eqn:globalprop}
\end{equation}
is obtained, where all the matrices involved are \emph{global}. The unknowns $\emf^{\localeq}_{\scatcontrib}$ from (\ref{eqn:localprop}) are now part of the unknown $\emf$ in (\ref{eqn:globalprop}) and appear in the positions corresponding to the primal edges of $\interface$.
Moreover, the terms $\emf_{\radcontrib}$ and $\mmf_{\radcontrib}$ are nonzero only in correspondence of the primal edges of $\interface$ and the dual edges of $\interface$ respectively. By introducing the terms due to the impedance boundary conditions (\ref{eqn:propagation-DGA-fb}) and plane wave excitation (\ref{eqn:complete-probl}), the full equation is obtained
\begin{equation}
    \mathbf{K}\emf + \admittancecontrib = -\mathbf{K}\emf_{\radcontrib} - i\omega\mmf_{\radcontrib} -2i\omega\bordermmf^-.
\end{equation}

\subsection{The equivalent antenna}
Given an antenna of arbitrary shape, radiated electric and magnetic fields can be computed in each point of space by means of tools like NEC or other, more advanced, simulators. In our case we used a dipole, for which the radiated electromagnetic field is known in closed form \cite{balanis}:
\begin{align}
    E_{\theta} &= \frac{i\eta I_0e^{-ikr}}{2\pi r}\left[\frac{\cos\left(\frac{kL}{2}\cos\theta\right) - cos\left(\frac{kL}{2}\right)}{\sin\theta}\right], \label{eqn:dipole-e}\\
    H_{\phi} &= \frac{E_{\theta}}{\eta} \label{eqn:dipole-h},
\end{align}
where $\eta$ is the impedance of free space, $I_0$ is the excitation current at the feedpoint (\ref{eqn:dipole-current}), $k$ is the propagation constant, $L$ is the dipole length and $r$ is the distance. In (\ref{eqn:dipole-e}) and (\ref{eqn:dipole-h}) it is assumed that the dipole lies along the $z$ axis of a cartesian reference system and its center is in the origin. On our simulator, arbitrary rotations of the virtual dipole are obtained by means of quaternions \cite{vince}. Once the field is known, it can be used to compute the values of the electromotive forces across the primal edges of the interface $\interface$ and the magnetomotive forces across its dual edges.
\section{Validation of the equivalent model} \label{sec:validation}
The diameter of the sphere representing the equivalent radiator and the average edge length of tetrahedra in its surroundings must be properly tuned, according to the operating frequency. To obtain reasonable performance, the diameter should be comparable with the wavelength $\lambda$ and the edge length should be comparable with $\lambda/10$. However, these are only rough indications: in our experiments, a single mesh tuned for the center frequency proved to be adequate between $\lambda = 3.3m$ and $\lambda = 0.77m$. To validate the equivalent model, a cube of side $l=5m$ was considered, while the radiator was represented by a sphere of radius $r=0.75m$ placed in the center of the cube. The boundary conditions on the six faces of the cube were set to \emph{impedance boundary condition} \cite{dga-admittance} with $Z = \sqrt{\mu_0/\epsilon_0}$. Electromagnetic wave propagation problem was solved and the field was evaluated at $r=2.5m$.

\begin{figure}[tb]
    \centering
    \includegraphics[width=0.7\linewidth]{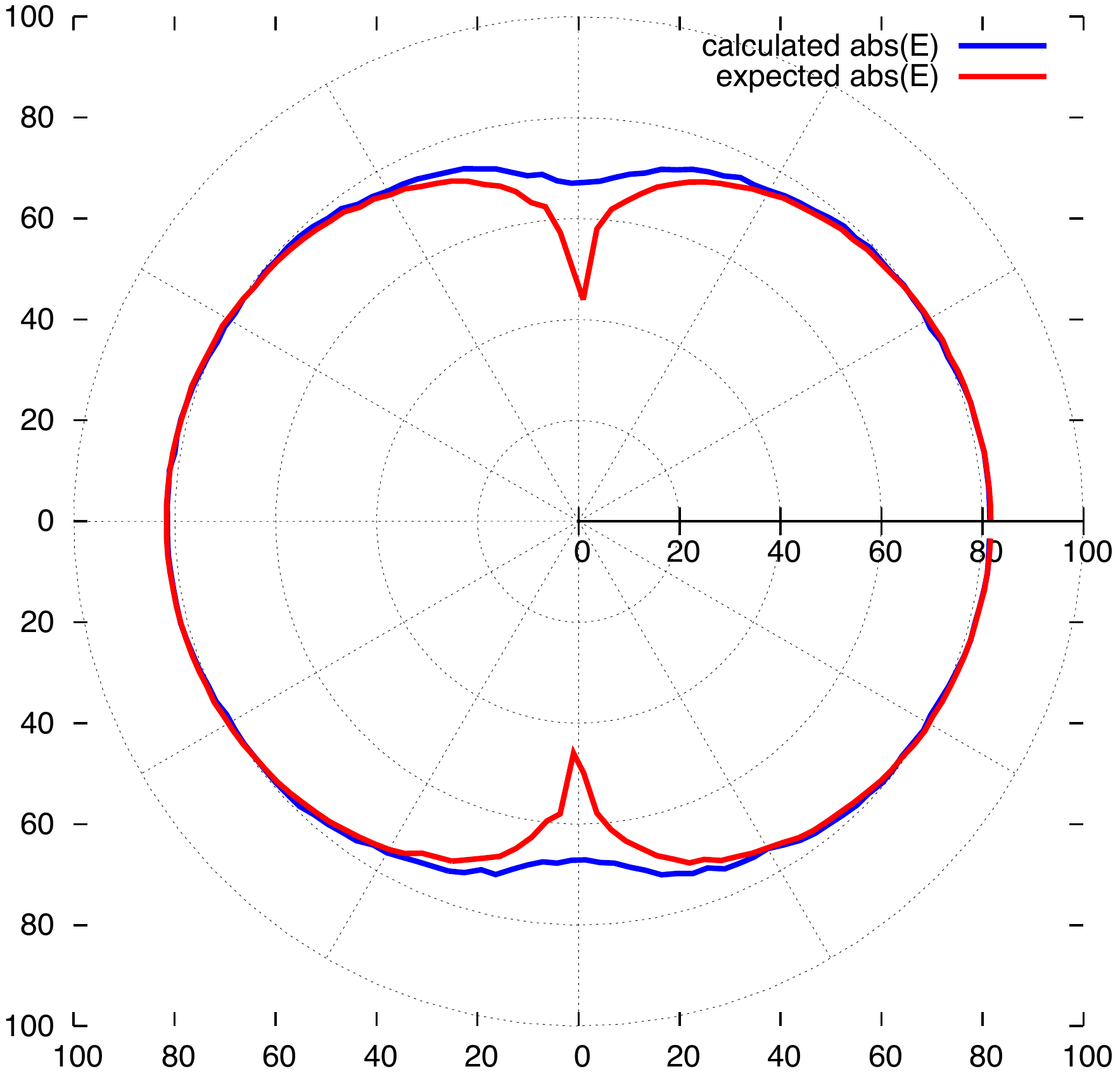}
    \caption{Comparison between the expected field and the field produced by the equivalent model at $f=230$ MHz. The field is expressed in $dB\mu V/m$.}
\end{figure}


\section{Modeling of an anechoic chamber}
Using the results developed in this work and in \cite{port-boundary-condition}, an entire anechoic chamber with a transmitter inside was simulated. The chamber, located at the Emilab EMC laboratory and manufactured by ETS-Lindgren, has dimensions (in~$x, y$~and~$z$) of $8.64m \times 5.6m \times 5.68m$. 
\subsection{Anechoic chamber walls}
The walls of the anechoic chamber were transformed to equivalent surfaces and treated as impedance boundary conditions using the methodology described in our previous work \cite{port-boundary-condition}. This step was done frequency by frequency and required 31 simulations. Cone and ferrite material parameters were provided by the manufacturer as tables of complex values of $\epsilon_r$ and $\mu_r$, in steps of 4 MHz. For the frequencies where tabulated data was not available, linear interpolation was used to calculate the value.
\subsection{Transmitter}
The transmitter was simulated by means of an equivalent radiator, modelled as a sphere meeting the requirements described in Section~\ref{sec:validation} and with a diameter of $r=0.75m$.

\section{Results}

\subsection{Numerical experiments}
The numerical simulations were made with the EMT code, our DGA workbench written in C++11. The first step was to calculate, for each frequency, the equivalent impedance parameters for the anechoic walls using the technique of \cite{port-boundary-condition}. Once the parameters were obtained, the entire anechoic chamber was described with a mesh of about 1.1 million elements. The simulations were performed on Debian/GNU Linux with kernel 3.14 running on a dual Xeon E5420 workstation with 32 GB of RAM, GCC 4.9.1 compiler and MKL PARDISO solver. The frequency sweep from 90 to 390 MHz for each antenna configuration required running a total of $31 \times 18 = 558$ simulations, which required slightly more than three days of runtime. A direct solver was employed, but we are investigating alternative solution techniques as multigrid.

\subsection{Comparison}
\begin{figure}[h]
    \centering
    \includegraphics[width=0.95\linewidth]{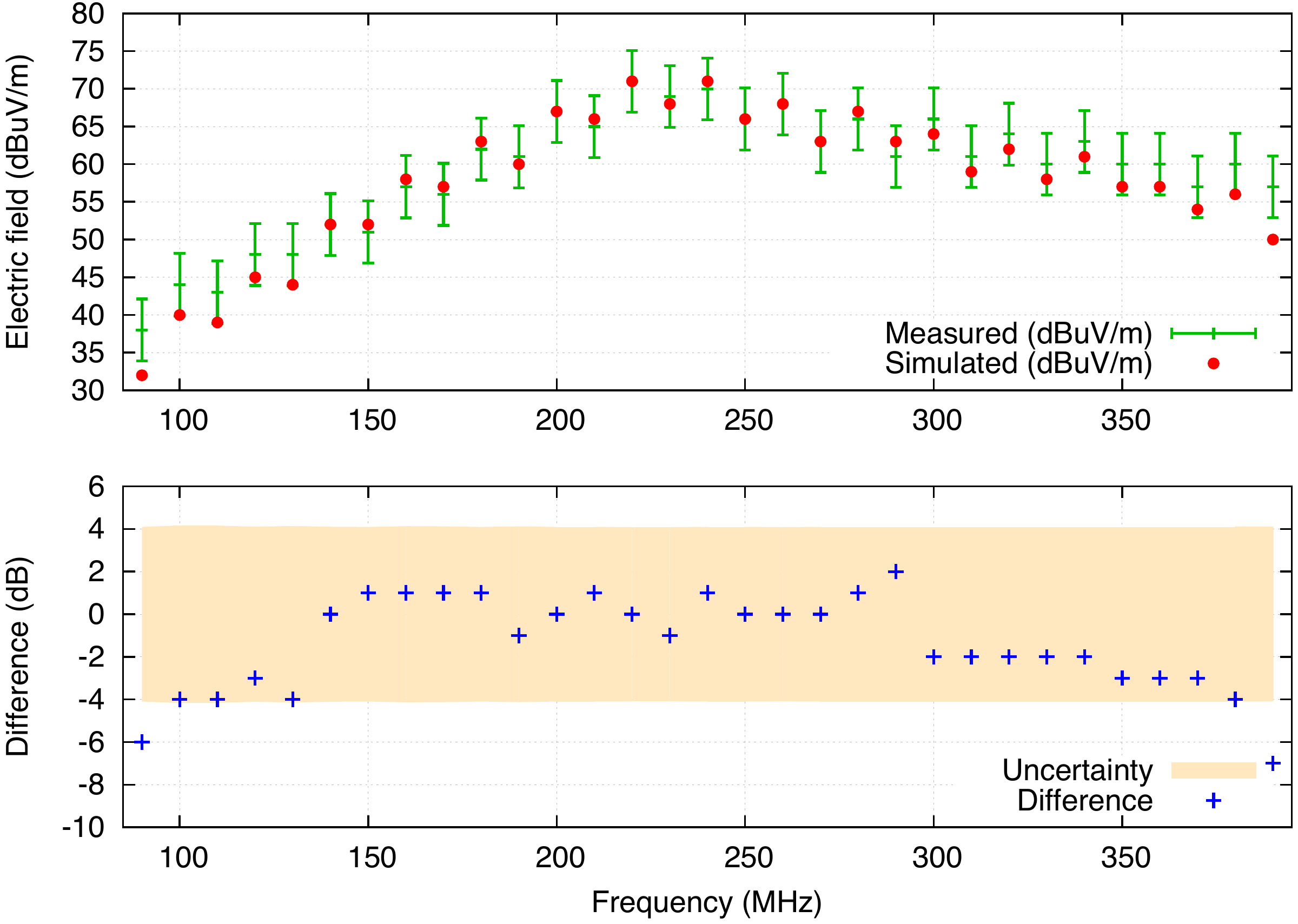}
    \caption{Comparison between computed field magnitude and measured field magnitude. Horizontal polarization is considered with both antennas at $h=1m$.}
    \label{fig:comp100h}
\end{figure}

\begin{figure}[h]
    \centering
    \includegraphics[width=0.95\linewidth]{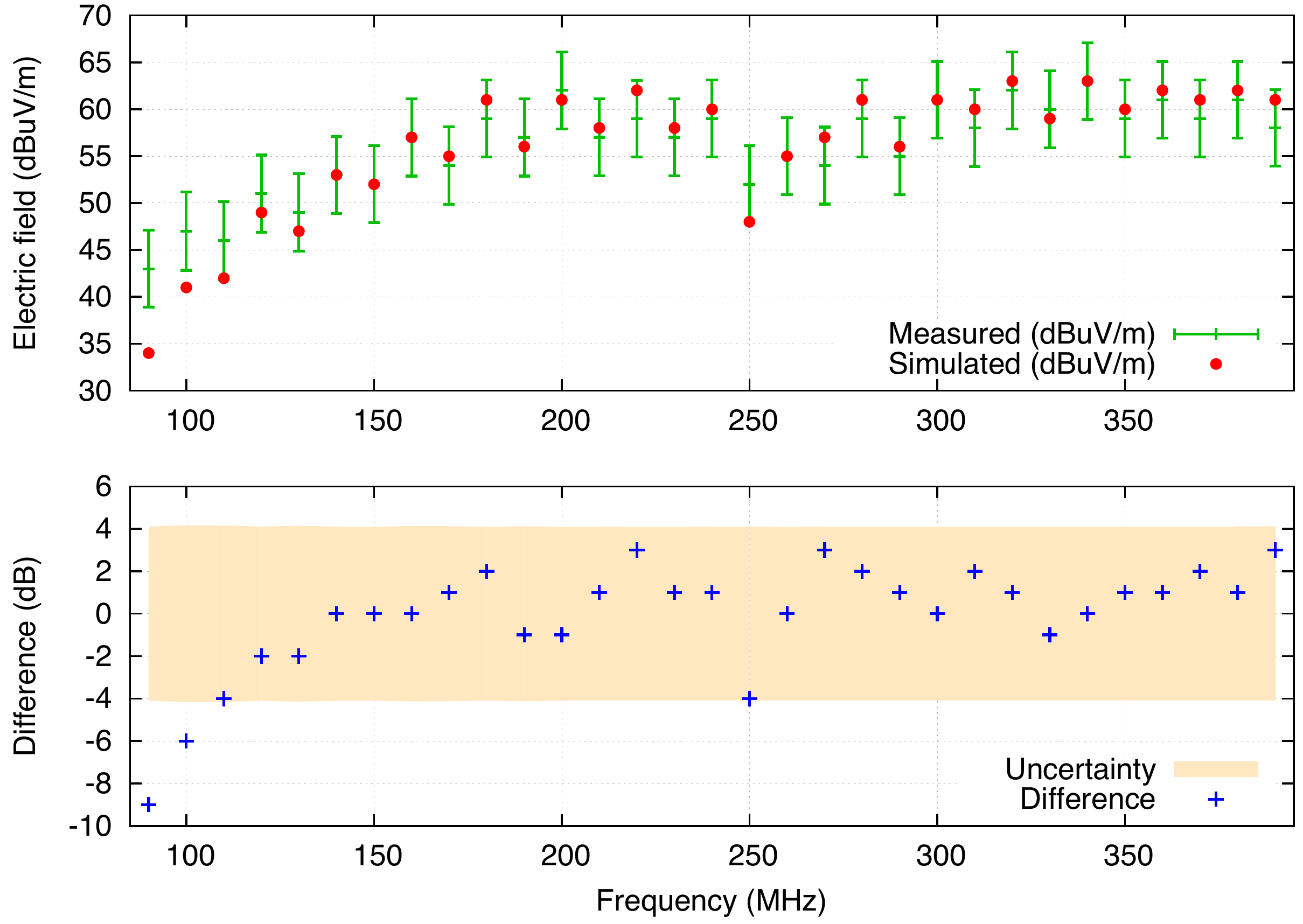}
    \caption{Comparison between computed field magnitude and measured field magnitude. Horizontal polarization is considered with both antennas at $h=1.5m$.}
    \label{fig:comp150h}
\end{figure}

\begin{figure}[h]
    \centering
    \includegraphics[width=0.95\linewidth]{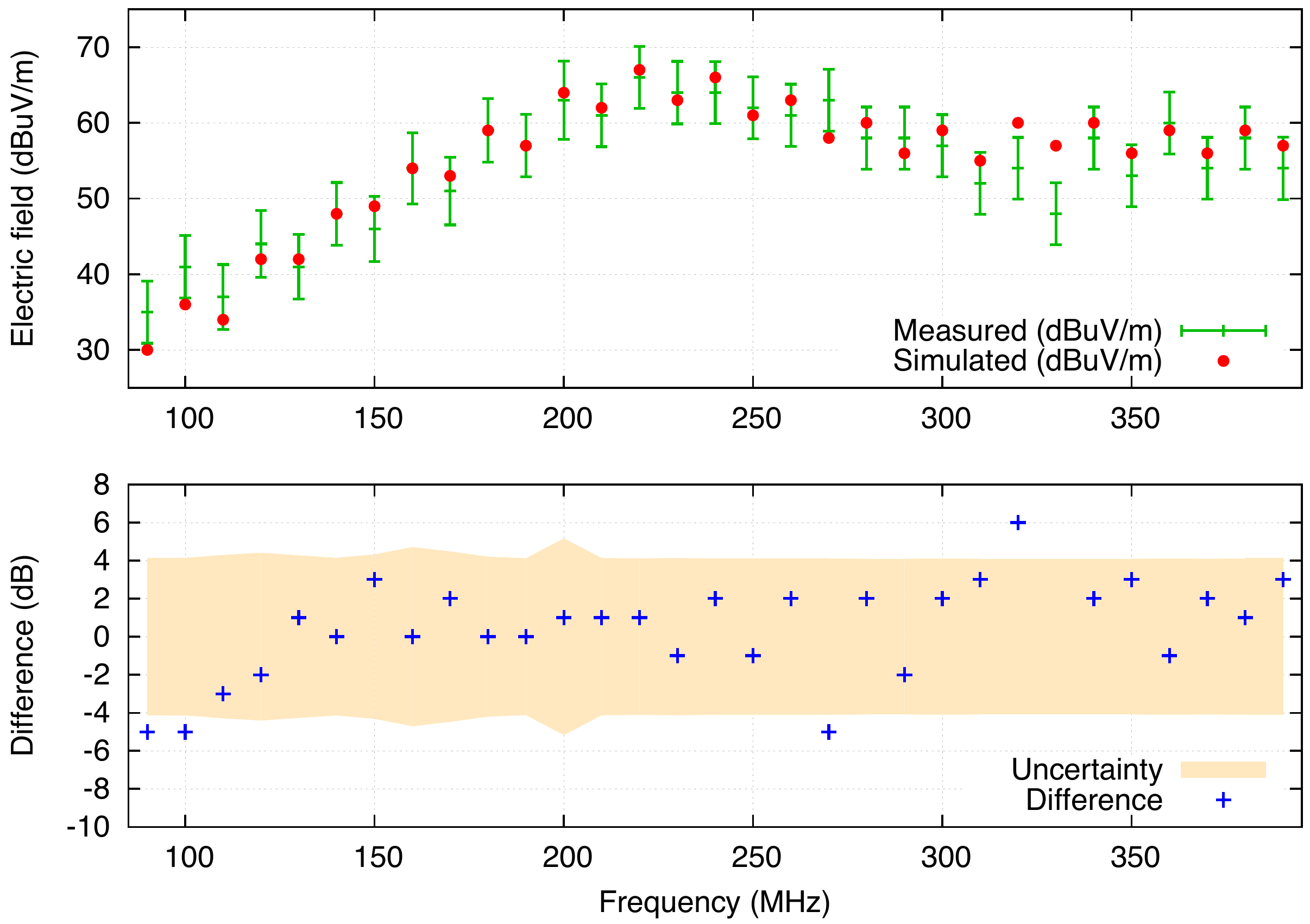}
    \caption{Comparison between computed field magnitude and measured field magnitude. Vertical polarization is considered with both antennas at $h=2m$.}
    \label{fig:comp200v}
\end{figure}

\begin{figure}[h]
    \centering
    \includegraphics[width=0.85\linewidth]{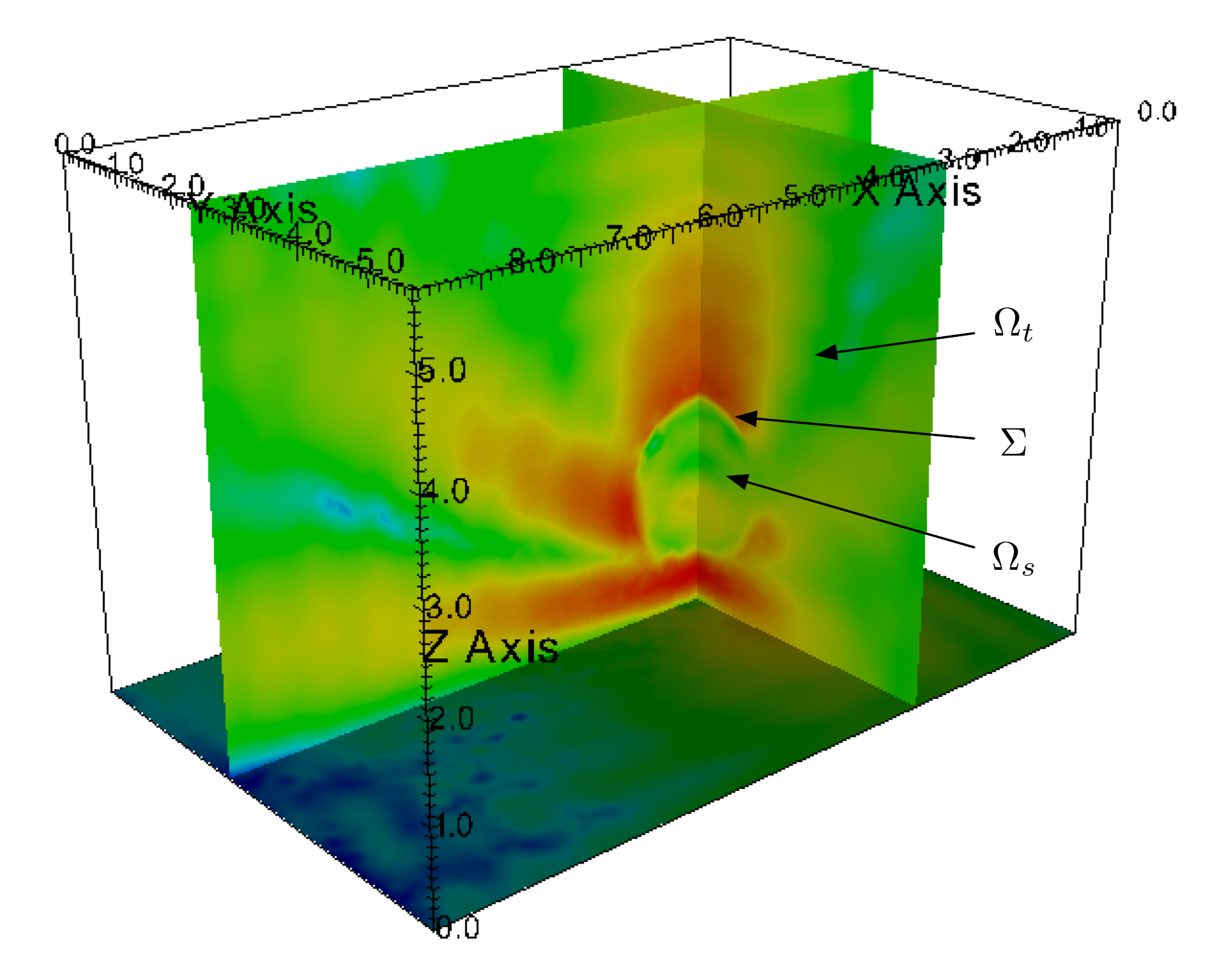}
    \caption{Sectional view of the electric field radiated by the equivalent dipole at $f=230$ MHz, horizontal polarization. The jump between the scattered field to the total field on the boundary of the sphere is clearly visible.}
    \label{fig:sim230-3d}
\end{figure}

\begin{figure}[h]
    \centering
    \includegraphics[width=0.85\linewidth]{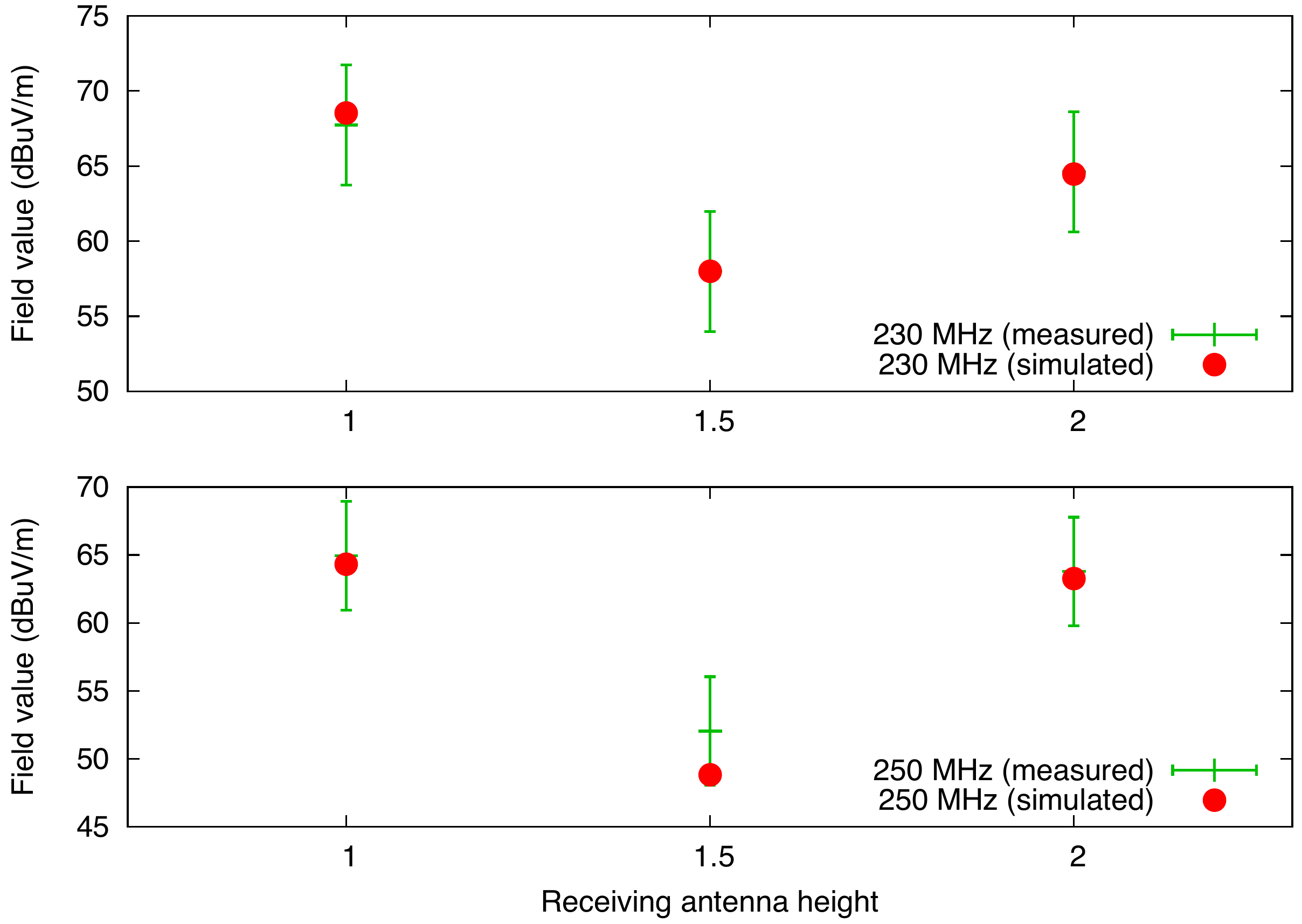}
    \caption{Simulated electric field and measured electric field are compared at different heights, $f=230$ MHz and $f=250$ MHz in horizontal polarization.}
    \label{fig:hole-230}
\end{figure}

From the comparison between experiments and simulations, we observed that our model is of remarkable accuracy.
In the following, some comparisons are shown. Figures  \ref{fig:comp100h}, \ref{fig:comp150h}, \ref{fig:comp200v} and \ref{fig:comp150v} contain the values of the measured field and the simulated field, their difference and the uncertainty band of the measurements. In almost all cases the simulation lies within the uncertainty band.
A further confirmation of the accuracy of the site model is given in Figs. \ref{fig:sim230-3d} and \ref{fig:hole-230}. Figure \ref{fig:sim230-3d} depicts a section of the field distribution inside the chamber. The transmitting dipole is at position $(2.565m, 2.3m, 1m)$ in horizontal polarization, while the receiving antenna was placed respectively at $(5.45m, 3.15m, 1m)$, $(5.45m, 3.15m, 1.5m)$ and $(5.45m, 3.15m, 2m)$. The distortion of the radiation pattern due to the conductive floor \cite{balanis} is clearly visible. Figure \ref{fig:hole-230} shows how the field varies with height at the three antenna positions and how the simulation correctly predicts the measured variation.
\subsection{On the discrepancies between simulations and measurements}
In Figs. \ref{fig:comp100h}, \ref{fig:comp150h} and \ref{fig:comp200v}, some simulation results fall outside the uncertainty band of the field measurement.
However, until now the antenna current used as input of the simulation was considered exact. But, as a measured quantity, the current has also an associated uncertainty which affects the result of the simulation. In the following we give an estimate of that uncertainty.
\begin{figure}[t]
    \centering
    \includegraphics[width=0.85\linewidth]{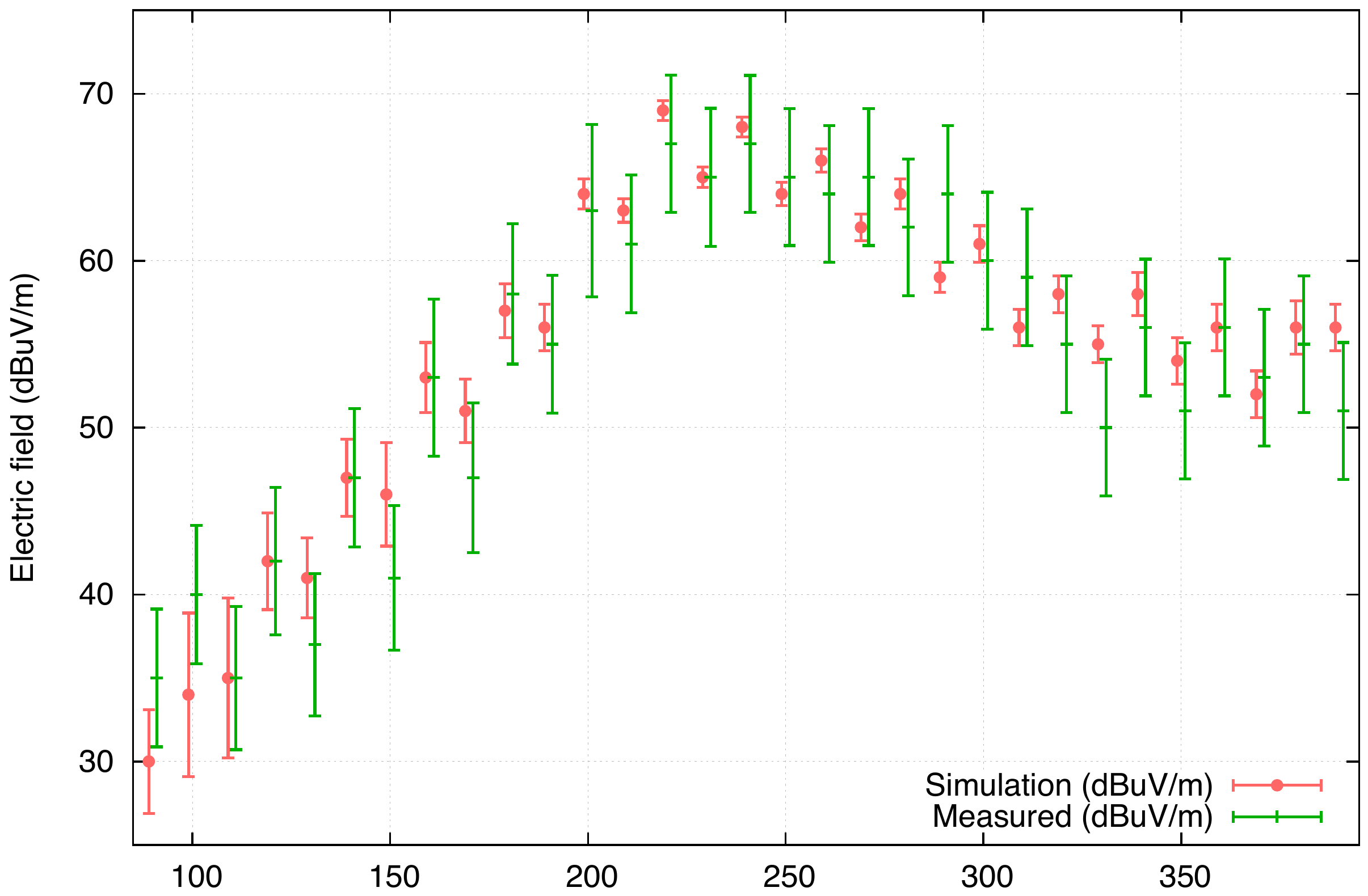}
    \caption{Comparison between computed field magnitude and measured field magnitude. Vertical polarization, both antennas at $h=1.5m$. The effect of the uncertainty on the antenna current, and thus in the simulation result, is shown.}
    \label{fig:comp150v}
\end{figure}
The electric field radiated by the dipole, calculated as in (\ref{eqn:dipole-e}), can be rewritten in logarithmic form as
\begin{align}
    E_{\theta,dB} = 20\log(I_0) + 20\log(T),
\end{align}
where $T$ accounts for all the multiplicative terms in (\ref{eqn:dipole-e}). Moreover, as already detailed, the current $I_0$ is calculated as
\begin{align}
    I_0 = \sqrt{\frac{P_{fwd} - P_{rev}}{Z_{ant}}}, \label{eqn:pwrcurrent}
\end{align}
where
\begin{align}
    Z_{ant} = Z_0\frac{1+\Gamma}{1-\Gamma}, \label{eqn:zantgamma}
\end{align}
with $\Gamma$ denoting the reflection coefficent. Therefore, using (\ref{eqn:pwrcurrent}) and (\ref{eqn:zantgamma}) the antenna current can be written as
\begin{align*}
    20\log(I_0) = 10\log(P_{fwd}) + 20\log(1-\Gamma) - 10\log(Z_0).
\end{align*}
The sensitivity coefficents can now be calculated \cite{cispr-16-4-2, gum1} as
\begin{align}
    c_{\Gamma} &= \frac{\partial(20\log I_0)}{\partial \Gamma} = -\frac{20}{1-\Gamma},\\
    c_p &= \frac{\partial(20\log I_0)}{\partial P_{fwd,dB}} = 1.
\end{align}
With these coefficents, the uncertainty of the radiated field can be calculated as
\begin{align}
    u(E_{\theta,dB}) = \sqrt{(c_p u_p)^2 + (c_{\Gamma} u_{\Gamma})^2},
\end{align}
where $u_p$ and $u_{\Gamma}$ have the values $0.5$ and $0.01$ respectively, and are derived from the instrument specifications. These considerations on the uncertainty of the input current give rise to a directly correlated uncertainty in the simulation output, depicted in Fig. \ref{fig:comp150v}. As expected, at frequencies where the mismatch between the dipole and the source is very high (dipole resonant frequency is 230 MHz), the uncertainty is also high because $P_{fwd}$ and $P_{rev}$ are of comparable magnitude.

\section{Conclusions}
Nowadays numerical simulation of anechoic chambers is a central tool in the study of their performance \cite{sim-large-anec,sim-field-unif}. However, the mathematical nature of the problem, the physical dimensions of the simulated sites and the frequency ranges pose non trivial challenges to the numerical treatment of the simulation problem. In this work we introduced a technique to describe anechoic walls and radiating elements as equivalent models, in order to reduce the degrees of freedom of the numerical problem and thus the required computational resources. Since important approximations were introduced, the validity of our equivalent models was tested against real measurements. That comparison proved the good quality and the usefulness of our technique.

\FloatBarrier

\end{document}